\documentclass[11pt]{article}
\usepackage{fixltx2e} 
\usepackage{cmap} 
\usepackage[T1]{fontenc}
\usepackage[utf8]{inputenc}
\usepackage{palatino}
\usepackage{multirow}
\usepackage{ifthen}
\usepackage{amsmath,amssymb}
\usepackage{longtable,ltcaption,array}
\newlength{\DUtablewidth} 

\makeatletter
\usepackage{fullpage}
\usepackage{moresize}
\usepackage{graphicx}

\makeatother
\makeatletter

\RequirePackage{color}

\definecolor{DUcolor0}{rgb}{1.00,0.00,0.00}

\expandafter\providecommand\csname DUrolec1\endcsname[1]{\textit{#1}}

\expandafter\providecommand\csname DUroles2\endcsname[1]{\textit{#1}}

\expandafter\providecommand\csname DUroles1\endcsname[1]{\textit{#1}}

\makeatother

\providecommand{\DUadmonition}[2][class-arg]{%
  \ifcsname DUadmonition#1\endcsname%
    \csname DUadmonition#1\endcsname{#2}%
  \else
    \begin{center}
      \fbox{\parbox{0.9\textwidth}{#2}}
    \end{center}
  \fi
}

\providecommand*{\DUtitle}[2][class-arg]{%
  \ifcsname DUtitle#1\endcsname%
    \csname DUtitle#1\endcsname{#2}%
  \else
    \smallskip\noindent\textbf{#2}\smallskip%
  \fi
}

\providecommand*{\DUtransition}[1][class-arg]{%
  \hspace*{\fill}\hrulefill\hspace*{\fill}
  \vskip 0.5\baselineskip
}

\ifthenelse{\isundefined{\hypersetup}}{
  \usepackage[colorlinks=true,linkcolor=blue,urlcolor=blue]{hyperref}
  \urlstyle{same} 
}{}
\hypersetup{
  pdftitle={Categories from scratch},
  pdfauthor={Raphael ‘kena’ Poss}
}

\title{\phantomsection%
  Categories from scratch%
  \label{categories-from-scratch}}
\author{Raphael ‘kena’ Poss}
\date{April 2014}

\begin{document}
\maketitle

\clearpage

\phantomsection\label{contents}
\pdfbookmark[1]{Contents}{contents}
\tableofcontents

\clearpage

\DUadmonition[note]{
\DUtitle[note]{Note}

The latest version of this document can be found online at
\url{http://science.raphael.poss.name/categories-from-scratch.html}.
Alternate formats:
\href{http://science.raphael.poss.name/categories-from-scratch.txt}{Source},
\href{http://science.raphael.poss.name/categories-from-scratch.pdf}{PDF}.
}

\section{Prologue%
  \label{prologue}%
}

The concept of \emph{category} from mathematics happens to be useful to
computer programmers in many ways. Unfortunately, all “good”
explanations of categories so far have been designed by
mathematicians, or at least theoreticians with a strong background in
mathematics, and this makes categories especially inscrutable to
external audiences.

More specifically, the common explanatory route to approach categories
is usually: “here is a formal specification of what a category is;
then look at these known things from maths and theoretical computer
science, and admire how they can be described using the notions of
category theory.” This approach is only successful if the audience can
fully understand a conceptual object using only its formal
specification.

In practice, \emph{quite a few people only adopt conceptual objects by
abstracting from two or more contexts where the concepts are
applicable,} instead. This is the road taken below: reconstruct the
abstractions from category theory using scratches of understanding from various
fields of computer engineering.

\section{Overview%
  \label{overview}%
}

The rest of this document is structured as follows:
\newcounter{listcnt0}
\begin{list}{\arabic{listcnt0}.}
{
\usecounter{listcnt0}
\setlength{\rightmargin}{\leftmargin}
}

\item introduction of example \hyperref[topics-of-study]{Topics of study}: unix process pipelines,
program statement sequences and signal processing circuits;

\item \hyperref[recollections]{Recollections} of some previous knowledge about each example;
highlight of interesting analogies between the examples;

\item \hyperref[identification-of-the-analogies]{Identification of the analogies} with existing concepts from category theory;

\item a quick preview of \hyperref[goodies-from-category-theory]{Goodies from category theory};

\item references to \hyperref[further-reading]{Further reading}.
\end{list}

\section{Topics of study%
  \label{topics-of-study}%
}
\begin{description}
\item[{\textbf{“Pipes”}}] \leavevmode 
Unix process pipelines: chains of Unix processes
linked by FIFOs.

\item[{\textbf{“Compilers”}}] \leavevmode 
Programs that transform programs.

\item[{\textbf{“Circuits”}}] \leavevmode 
Signal processing circuits: circuits in the real world with
physical connectors for input and output of electrical signals.

\end{description}

Many introductory texts about category theory use another group of
examples: type systems for programming languages, matrices from linear
algebra and sometimes directed graphs. I did not choose them here as I
believe they are already too abstract for many computing engineers.
Instead, I believe these examples should only be mentioned as
additional instances of categories after the concept has been
properly recognized.

\section{Recollections%
  \label{recollections}%
}

\subsection{Background terminology%
  \label{background-terminology}%
}

\setlength{\DUtablewidth}{\linewidth}
\begin{longtable*}[c]{p{0.315\DUtablewidth}p{0.315\DUtablewidth}p{0.315\DUtablewidth}}

\textbf{Pipes:}

In a Unix system, processing activities are organized in
\emph{processes}.  Nearly all \emph{communication} between
processes and with I/O devices is organized via file
descriptors: to real files on disk, to terminals with the
user, but also FIFO buffers between
processes. Communication can be monitored directly on
terminals, or by looking at the contents of files on disk, or
interleaving the program \texttt{tee} between programs chained by
FIFOs.
 & 
\textbf{Compilers:}

A compiler is a program that takes another program as input,
and transforms it to produce another program as output. The
set of valid inputs for a compiler is its \emph{input language}, and
the set of possible outputs is its \emph{output language}. For
example, a “C compiler” accepts C code as input and produces
assembly code as output (for some specific ISA).
 & 
\textbf{Circuits:}

An electronic circuit is usually recognized when one sees a
plastic enclosure with some metal bits sticking out. \emph{Signal
processors} are a specific type of circuit, with a notion of
“input” and “output” connectors: For example, a hi-fi amplifier
has distinct pins for the “audio source” and one or more
“speaker” output connectors. When \emph{plugged in} to an active
input signal, the circuit is itself activated and starts
driving its output pins. The output signal can be exploited
by further plugging to other circuits, or output devices. It
can also be measured, for example using an oscilloscope.
 \\
\end{longtable*}

\subsection{Observability%
  \label{observability}%
}

Each topic has two \textquotedbl{}groups\textquotedbl{} of \emph{things}: things that can be \emph{observed}
from the “outside”, and things that appear (mostly) as \emph{black boxes} from
the outside but are “connected” between the things in the first group:

\setlength{\DUtablewidth}{\linewidth}
\begin{longtable*}[c]{p{0.131\DUtablewidth}p{0.267\DUtablewidth}p{0.276\DUtablewidth}p{0.276\DUtablewidth}}

Example:
 & 
\textbf{Pipes}
 & 
\textbf{Compilers}
 & 
\textbf{Circuits}
 \\

Observables:
 & 
The data streams
going through file
descriptors.
 & 
The set of all possible
programs written in the
input languages or
generated in the
output language.
 & 
The electric signals over
time.
 \\

Black boxes:
 & 
The running
processes with their
internal state
(program counter,
stack, heap, etc.).
 & 
The program transformation
algorithms.
 & 
The circuits themselves.
 \\
\end{longtable*}

\subsection{Grouping things together%
  \label{grouping-things-together}%
}

Each topic provides a mechanism to plug the \textquotedbl{}black box\textquotedbl{} things together:

\setlength{\DUtablewidth}{\linewidth}
\begin{longtable*}[c]{p{0.315\DUtablewidth}p{0.315\DUtablewidth}p{0.315\DUtablewidth}}

\textbf{Pipes:}

Given the two commands \textquotedbl{}\texttt{grep -v foo}\textquotedbl{} and
\textquotedbl{}\texttt{grep -v bar}\textquotedbl{}, one can write the command \textquotedbl{}\texttt{grep -v foo |
grep -v bar}\textquotedbl{}, which combines the two behaviors.
 & 
\textbf{Compilers:}

Given a compiler from C to assembly text, and a compiler from assembly text
to machine code, one can combine them (eg. by means of a script) to
create a compiler from C to machine code.
 & 
\textbf{Circuits:}

Given the a DVI-VGA adapter and a VGA-SVideo adapter, it is
possible to plug them together to form a DVI-SVideo adapter.
 \\
\end{longtable*}

\subsection{Composition semantics%
  \label{composition-semantics}%
}

Each topic has a notion of \textquotedbl{}good\textquotedbl{}  compositions that \textquotedbl{}make sense\textquotedbl{}, and
\textquotedbl{}bad\textquotedbl{} compositions that nonsensical and not expected to \textquotedbl{}work
properly\textquotedbl{}:

\setlength{\DUtablewidth}{\linewidth}
\begin{longtable*}[c]{p{0.315\DUtablewidth}p{0.315\DUtablewidth}p{0.315\DUtablewidth}}

\textbf{Pipes:}

Piping \texttt{ls} with \texttt{grep foo} is sensical.
 & 
\textbf{Compilers:}

Connecting the output of a C-to-assembly compiler
to the input of a assembly-to-code compiler is sensical.
 & 
\textbf{Circuits:}

Plugging a USB-serial adapter to a DB9-DB25 serial
adapter (with a 9-pin interface between them) is sensical.
 \\

Piping \texttt{ls} with \texttt{gv} (PostScript viewer) is nonsensical.
 & 
Connecting the output of a C-to-C compiler to the input
of an assembly-to-code compiler is nonsensical.
 & 
Plugging a USB-serial adapter to an EGA display (physically
possible as they share the same DB9 connector) is nonsensical.
 \\
\end{longtable*}

To identify \textquotedbl{}good\textquotedbl{} from \textquotedbl{}bad\textquotedbl{} compositions, each topic places a large
emphasis on the notion of \emph{interface}:

\setlength{\DUtablewidth}{\linewidth}
\begin{longtable*}[c]{p{0.315\DUtablewidth}p{0.315\DUtablewidth}p{0.315\DUtablewidth}}

\textbf{Pipes:}

Both \texttt{ls} and \texttt{grep} operate on plain text streams, which
is why they compose well with pipes. In contrast \texttt{gv} expects
PostScript as input, which \texttt{ls} cannot produce.
 & 
\textbf{Compilers:}

Each compiler has a notion of \emph{input language} for the set of accepted
input programs and \emph{output language} for the set of possible outputs.
The languages must match when composing the compilers together.
 & 
\textbf{Circuits:}

The USB-serial adapter and serial-serial adapter plug well
together because they both use the same standard (RS232)
signalling protocol at the interface.
 \\
\end{longtable*}

Remarkably, users \emph{understand} or conceptualize interfaces, despite
the fact they are not always defined explicitly beforehand.

\subsection{Neutral behaviors%
  \label{neutral-behaviors}%
}

Each topic has some instances of black box things that \textquotedbl{}do nothing\textquotedbl{},
ie have a \emph{neutral} behavior:

\setlength{\DUtablewidth}{\linewidth}
\begin{longtable*}[c]{p{0.315\DUtablewidth}p{0.315\DUtablewidth}p{0.315\DUtablewidth}}

\textbf{Pipes:}

when the input and output data streams are the same (byte-wise).
 & 
\textbf{Compilers:}

when the input and output languages are the same.
 & 
\textbf{Circuits:}

when the output signals are measured the same as the input.
 \\
\end{longtable*}

The \emph{knowledge} of whether a black box is neutral can be gained in either or three ways.

Either it is known to be neutral \emph{by construction}, because the
specification is available for scrutinity and can be proven to define
a neutral behavior:

\setlength{\DUtablewidth}{\linewidth}
\begin{longtable*}[c]{p{0.315\DUtablewidth}p{0.315\DUtablewidth}p{0.315\DUtablewidth}}

\textbf{Pipes:}

The program source code contains a loop that iteratively
inputs a byte and outputs the same byte, stopping at
end-of-stream.
 & 
\textbf{Compilers:}

The transformation algorithm does not change the semantics
of the program.
 & 
\textbf{Circuits:}

The blue print defines direct links between the input and
output pins.
 \\
\end{longtable*}

Or, the knowledge is provided \emph{externally}, eg. by fiat:

\setlength{\DUtablewidth}{\linewidth}
\begin{longtable*}[c]{p{0.315\DUtablewidth}p{0.315\DUtablewidth}p{0.315\DUtablewidth}}

\textbf{Pipes:}

The manual page for a command specifies that the process
will replicate its input to its output unchanged.
 & 
\textbf{Compilers:}

The documentation says it is a “source-to-source” compiler,
or explicitly indicates that its input and output language are the same.
 & 
\textbf{Circuits:}

The manufacturer guarantees that the circuit is fully
pass-through.
 \\
\end{longtable*}

Or, it is \emph{discovered}: to find out whether a black box thing A is
neutral, assuming an observer has access to a pre-existing, valid
\textquotedbl{}observable thing\textquotedbl{} that can be fed to A, then the observer can deduce
the behavior of A is \textquotedbl{}neutral\textquotedbl{} if the observable as a
result of A's activity is the same as the original observable:

\setlength{\DUtablewidth}{\linewidth}
\begin{longtable*}[c]{p{0.315\DUtablewidth}p{0.315\DUtablewidth}p{0.315\DUtablewidth}}

\textbf{Pipes:}

An unknown program \texttt{xxx} is known \emph{a priori} to only read
from its standard input and write to its standard output.  So
one can run the command \textquotedbl{}\texttt{xxx <iN >oN}\textquotedbl{} for various input
files \texttt{iN} and compare whether the contents of each file
\texttt{oN} are equal to the corresponding \texttt{iN}. If so, \texttt{xxx}
appears to be \textquotedbl{}neutral\textquotedbl{}.
 & 
\textbf{Compilers:}

The input language is known, but not the output language.  So
one can generate some random but valid input programs and feed
them to the compiler. If the output programs are also valid
input programs, then original program piece appears to be
\textquotedbl{}neutral\textquotedbl{}.
 & 
\textbf{Circuits:}

A signal processor appears to have the same number of input and
output pins, and its input signal specification is known \emph{a
priori}. So one can use various valid input signals, feed them
to the circuit, and measure the output. If the output signals
measures the same as the input every time, the circuit appears
to be \textquotedbl{}neutral\textquotedbl{}.
 \\
\end{longtable*}

Once a black box thing N is known to be neutral, then its composition
\textquotedbl{}left\textquotedbl{} and \textquotedbl{}right\textquotedbl{} with another black box thing A can be assumed to
have the same behavior as A on its own:

\setlength{\DUtablewidth}{\linewidth}
\begin{longtable*}[c]{p{0.315\DUtablewidth}p{0.315\DUtablewidth}p{0.315\DUtablewidth}}

\textbf{Pipes:}

Both \textquotedbl{}\texttt{xxx | grep foo}\textquotedbl{} and \textquotedbl{}\texttt{grep foo | xxx}\textquotedbl{} can be
assumed to behave like \texttt{grep foo} once \texttt{xxx} is known to be
neutral.
 & 
\textbf{Compilers:}

A compiler built by composing a neutral compiler N either before or after
another compiler C will have the same input and output languages as C.
 & 
\textbf{Circuits:}

Plugging a neutral circuit on either the input or output side
of another circuit A will process signals as A alone would.
 \\
\end{longtable*}

\subsection{Associativity%
  \label{associativity}%
}

When one composes three black box things A, B and C together (assuming
the compositions are point-wise sensical), the \emph{order} in which the
composition is realized does not change the behavior:

\setlength{\DUtablewidth}{\linewidth}
\begin{longtable*}[c]{p{0.315\DUtablewidth}p{0.315\DUtablewidth}p{0.315\DUtablewidth}}

\textbf{Pipes:}

The commands \textquotedbl{}\texttt{(ls | grep -v foo) | grep -v bar}\textquotedbl{} and \textquotedbl{}\texttt{ls |
(grep -v foo | grep -v bar)}\textquotedbl{} have equivalent behavior on
their data streams.
 & 
\textbf{Compilers:}

If a script A invokes a Scheme-to-C compiler C1 followed by a C-to-assembly
compiler C2, and another script B invokes A and then an
assembly-to-code compiler C3, then B has the same input and output
language as a script C that calls C1 then D, where D calls C2 then C3.
 & 
\textbf{Circuits:}

Whether a DVI-VGA adapter is plugged into a VGA-SVideo adapter,
and then the result is plugged
into a TV screen with SVideo input, or if a VGA-SVideo adapter
is first plugged into the TV,
and then plugged to a DVI-VGA adapter, both resulting circuits
are working TVs from a DVI input
signal.
 \\
\end{longtable*}

\section{Identification of the analogies%
  \label{identification-of-the-analogies}%
}

The previous section has introduced the following key concepts:
\begin{itemize}

\item things that can be \textquotedbl{}observed\textquotedbl{} from the outside, and \textquotedbl{}black box\textquotedbl{}
things connected between observables;

\item neutral black box things that preserve behavior;

\item associative composition of black box things.

\end{itemize}

These are the concepts manipulated in category theory.

\subsection{Objects and morphisms%
  \label{objects-and-morphisms}%
}

First, the observables are named \textbf{objects}. The black boxes things are named \textbf{morphisms} or \textbf{arrows}.

\setlength{\DUtablewidth}{\linewidth}
\begin{longtable*}[c]{p{0.139\DUtablewidth}p{0.263\DUtablewidth}p{0.274\DUtablewidth}p{0.274\DUtablewidth}}

Example:
 & 
\textbf{Pipes}
 & 
\textbf{Compilers}
 & 
\textbf{Circuits}
 \\

Objects:
 & 
Data streams.
 & 
Languages.
 & 
Signals.
 \\

Morphisms:
 & 
Processes.
 & 
Compilers.
 & 
Circuits.
 \\
\end{longtable*}

\subsection{Modeling: be careful about equivalences%
  \label{modeling-be-careful-about-equivalences}%
}

Categories are defined over \emph{mathematical sets} of objects and morphisms.
Sets are different from simple “collections” (or “bag”) of things from the real world:
all their elements are distinct,  according to some \emph{equivalence} relation.

So in order to talk about categories over things from the real world, we
must first choose how to define the mathematical sets. This choice
is called a \emph{model} and multiple models are possible for the same collection of things.

The most focus should be given to the set of morphisms. The set of objects
is simply derived from it once the morphisms are properly identified. For example,
in our pipes example, considering what happens to data streams. What
does it mean for two processes to be equal or different?

We can choose for example “data stream equality”. By this standard, two
processes that filter out lines containing the text “foo” over any
data stream are the “same thing.”  So “\texttt{sed '/foo/d'}” and
“\texttt{grep -v foo}” are the same morphism.

If we choose this definition for morphisms, then the objects are not
individual files (or time-particular datastreams over FIFOs), but
rather entire classes of all possible data streams that compare
equal to each other byte by byte. For example, a stream that
delivers “helloworld” in one go is the same stream as another that
delivers “hello” and then “world” 5 seconds later.

Another possible choice for a definition is \textquotedbl{}physical equality\textquotedbl{}. By this standard, two processes
that run at different times or in different physical regions of the
system are distinct, even if they perform the same task. So two
processes run from the same command (eg \texttt{cat}) at different times
end up as different morphisms in the set.

If we choose this definition, then the mathematical objects are not
only data streams, but also where and when the bytes are physically
encoded. So two streams that deliver “helloworld” in different
places/times are distinct objects.

The rule of thumb while choosing a definition is the
following: if one wants to talk about categories over sets of objects
and morphisms that are \emph{already} mathematically defined, then all is
well. If one wants to use category theory over things that are not
yet mathematical, be careful to explain \emph{clearly and explicitly} how they are modeled
using mathematical sets, and which equivalence relation is used.

For the next sections, we use the following definitions:

\setlength{\DUtablewidth}{\linewidth}
\begin{longtable*}[c]{p{0.315\DUtablewidth}p{0.315\DUtablewidth}p{0.315\DUtablewidth}}

\textbf{Pipes:}

We use “data stream equality” as defined above. With this, the
commands “\texttt{sed '/foo/d'}” and “\texttt{grep -v foo}” define the same
morphism; so do “\texttt{tr x y}” and “\texttt{sed s/x/y/g}”.

With this definition, each morphism may have multiple names
(different commands to define it). This is ok.
 & 
\textbf{Compilers:}

We use “program equality”: two compilers are the same morphism
if they produce the same output program from the same input
program. By this definition, two different C-to-assembly
compilers (eg. \texttt{gcc} and \texttt{clang}) are different morphisms,
but a compiler defined by a script combining \texttt{cpp} with
\texttt{gcc} is the same morphism as \texttt{gcc} on its own.

Again, with this definition, each morphism may have multiple names.
 & 
\textbf{Circuits:}

We use “interface and protocol equality”. With this, both the
\href{http://www.serialgear.com/2-Port-Serial-USB-USBG-232FT-1.html}{USBGEAR/USBG-232FT-1} and \href{http://www.serialgear.com/1-Port-Serial-USB-U232-P9.html}{MCT/U232-P9} are the same
morphism, as they both interface USB to RS232. However, a
USB-to-EGA adapter would not be the same morphism, because it
uses a different signal protocol even though the interface
(DB9) is the same.
 \\
\end{longtable*}

\subsection{Arrow notation%
  \label{arrow-notation}%
}

The \textbf{arrow notation} \emph{describes} a morphism: “$A\rightarrow B$”
is a description for a morphism from object \emph{A} to object \emph{B}.  The
two sides of a morphism can be named “input” and “output”, but are
usually named “source” and “target”.

There may be multiple morphisms between
any two objects, so the arrow notation does not identify
a particular morphism; instead, it can be seen as an “interface” or “type” for the
set of all morphisms between the designated objects:

\setlength{\DUtablewidth}{\linewidth}
\begin{longtable*}[c]{p{0.315\DUtablewidth}p{0.315\DUtablewidth}p{0.315\DUtablewidth}}

\textbf{Pipes:}

Both the processes resulting from running “\texttt{tr x y <f1
>f2}” and “\texttt{tr yz <f1 >f2}” are different morphisms, and both can be
described by the arrow “$\texttt{f1}\rightarrow\texttt{f2}$”.
 & 
\textbf{Compilers:}

Both the \texttt{gcc} and \texttt{clang} programs are morphisms, and both
can be described by the arrow “$\text{C}\rightarrow\text{assembly}$”.
 & 
\textbf{Circuits:}

Both the \href{http://www.serialgear.com/2-Port-Serial-USB-USBG-232FT-1.html}{USBGEAR/USBG-232FT-1} and \href{http://www.serialgear.com/1-Port-Serial-USB-U232-P9.html}{MCT/U232-P9} are morphisms,
and both can be described by the arrow “$\text{USB}\rightarrow\text{RS232}$”.
 \\
\end{longtable*}

\subsection{Composition%
  \label{composition}%
}

If two morphisms have the same intermediary object, it is possible to
compose them together (cf. \hyperref[composition-semantics]{Composition semantics} above).
This is abstracted by an \textbf{composition operator} noted
\textquotedbl{}$\cdot$\textquotedbl{}: given two compatible morphisms \emph{f} and \emph{g}, \textquotedbl{}$g\cdot f$\textquotedbl{}
designates their composition, ie. $(g\cdot f)(x) = g(f(x))$.

By construction, if \emph{f} can be described by $A\rightarrow B$ and
\emph{g} by $B\rightarrow C$, then $g\cdot f$ can be described
by $A\rightarrow C$.

Composition is associative: for any \emph{f}, \emph{g}, \emph{h}, $(f\cdot g)\cdot h = f \cdot (g\cdot h)$.

\subsection{Identity%
  \label{identity}%
}

For any object, there must exist at least one morphism that keeps the
object unchanged.  Each such \textbf{identity} morphism for an object \emph{x}
is called “$\text{id}_x$” (sometimes also “$\mathbf{1}_x$”) and can be described as
$x\rightarrow x$; it must satisfy the following property: for
every morphism $f : A\rightarrow B$, $\text{id}_{B}\cdot f
= f = f\cdot\text{id}_A$.

So there must be at least as many identity morphisms as there are objects.

For categories defined by modeling over concrete things, it may be
necessary to extend the mathematical set of morphisms with
“theoretical” identity morphisms, when there are no concrete
identities.

For example, it is not possible to build a concrete identity circuit:
any one-to-one pairing of physical input and output connectors with
direct wires between them is bound to introduce noise in the signal
due to the physical distance. However, the mathematical set modeling
circuits can be naturally extended to include “virtual” identity
circuits that preserve signals unchanged.

Hopefully, with many categories the identity morphisms can be
concretely constructed in the application domain:

\setlength{\DUtablewidth}{\linewidth}
\begin{longtable*}[c]{p{0.315\DUtablewidth}p{0.315\DUtablewidth}p{0.315\DUtablewidth}}

\textbf{Pipes:}

For any data stream, the morphism defined from
the commands “\texttt{cat}” or “\texttt{grep '.*'}” is an identity.
 & 
\textbf{Compilers:}

For the C language, the preprocessor (\texttt{cpp}) is an identity.
 & 
\textbf{Circuits:}

see above.
 \\
\end{longtable*}

\subsection{Definition of a category%
  \label{definition-of-a-category}%
}

A \textbf{category} is an \emph{algebraic structure} formed over:
\begin{itemize}

\item a (mathematical) set of objects,

\item a (mathematical) set of morphisms over these objects containing
at least one identity morphism for each object,

\item an associative composition operator.

\end{itemize}

\setlength{\DUtablewidth}{\linewidth}
\begin{longtable*}[c]{p{0.148\DUtablewidth}p{0.260\DUtablewidth}p{0.271\DUtablewidth}p{0.271\DUtablewidth}}

Category:
 & 
\textbf{Pipes}
 & 
\textbf{Compilers}
 & 
\textbf{Circuits}
 \\

Objects:
 & 
Data streams.
 & 
Languages.
 & 
Signals.
 \\

Morphisms:
 & 
Stream transformers.
 & 
Compilers.
 & 
Models of circuits.
 \\

Identities:
 & 
Running \texttt{cat} or
similar pass-through
commands.
 & 
\texttt{cpp} for C, in
general \texttt{cat} for any
langage.
 & 
Virtual pass-through
circuits.
 \\

Composition:
 & 
Chaining processes via
FIFO buffers, eg
by running them with
the pipe operator in
commands.
 & 
Creating a script that
invokes two existing
compilers, applying the
2nd on the output of the
1st.
 & 
Plugging the circuits
together.
 \\
\end{longtable*}

\section{Goodies from category theory%
  \label{goodies-from-category-theory}%
}

\subsection{Unicity of the identity%
  \label{unicity-of-the-identity}%
}

Although the \emph{existence} of identity morphisms is a prerequisite to
form a category (axiomatic), it is possible to prove within category
theory that each identity is \emph{unique}.

You can do this as follows.

Suppose you have two morphisms $\text{id1}_x$ and
$\text{id2}_x$ that preserve object \emph{x} and satisfy the axiomatic identity
properties:
\setcounter{listcnt0}{0}
\begin{list}{\arabic{listcnt0}.}
{
\usecounter{listcnt0}
\setlength{\rightmargin}{\leftmargin}
}

\item for every $f : y\rightarrow x$, $\text{id1}_x\cdot f = \text{id2}_x\cdot f = f$; and

\item for every $g : x\rightarrow y$, $g\cdot\text{id1}_x = g\cdot\text{id2}_x = g$.
\end{list}

In equation \#1, replace \emph{f} by $\text{id2}_x$, and you find that $\text{id1}_x\cdot \text{id2}_x = \text{id2}_x$.

In equation \#2, replace \emph{g} by $\text{id1}_x$, and you find that $\text{id1}_x\cdot \text{id2}_x = \text{id1}_x$.

Since both left-hand sides are equal, you have proven that
$\text{id2}_x = \text{id1}_x$. $\blacksquare$

\textbf{What this means in practice:} if you can construct/define two
morphisms in a category and prove they satisfy the identity laws, then
you have proven they are the same morphism. In our examples, that
means identity commands (pipes), compilers or circuits become
interchangeable with regards to their properties in category theory. This can
be used to simplify formulas that use complex morphisms into simpler ones.

\subsection{Invertibility and isomorphisms%
  \label{invertibility-and-isomorphisms}%
}

The general notion of invertibility for a morphism can be expressed
purely in the vocabulary of category theory:

$f : A\rightarrow B$ is invertible if there exists $g : B\rightarrow A$ such
that $g\cdot f : \text{id}_{A}$ and $f\cdot g : \text{id}_B$.

Invertible morphisms are also called \textbf{isomorphisms}.

By extension, two \emph{objects} \emph{A} and \emph{B} are \textbf{isomorphic} if there exists
at least one isomorphism described by $A\rightarrow B$.

\subsection{Duality%
  \label{duality}%
}

For any category $C$, it is possible to define mathematically
another category $C^{op}$ where the source and target of every
morphism are interchanged. This is called the “opposite category” or \textbf{dual category} of $C$.

Duality is “invertible”: $(C^{op})^{op}$ is the same category as $C$.

The dual category of a category $C$ of concrete objects may be
purely abstract, ie. without concrete representations for morphisms in
the application domain of $C$, for example if the morphisms in
$C$ are not invertible.

Nevertheless, duality serves an important purpose: say, you have
demonstrated a property that holds within a category $C$,
which you can express within the language of category theory using
a formula $\sigma$ (some text string).

If you then replace all occurences of “source” by “target” and
vice-versa in $\sigma$, and all occurences of $f\cdot g$
by $g\cdot f$, you obtain a new formula $\sigma^{op}$. By
construction, this formula is true in $C^{op}$. It is said to be
the \textbf{dual property} of $\sigma$.
Conversely, if you know a property $\sigma$ to be true in $C^{op}$, then
$\sigma^{op}$ will be true in $C$ as well.

\textbf{Why this is useful:} many important/useful results and properties
of mathematics come in pairs that are expressed using “symmetric”
formulas, which are dual in category theory.  For example,
\textbf{monomorphisms} and \textbf{epimorphisms} are morphisms for which
different properties hold, but their definitions are dual. From this,
if one can derive another unrelated property $\phi$ that relies on the fact a
morphism is monomorphic, then thanks to duality, automatically the
dual property $\phi^{op}$ is also proven over epimorphisms.
Thanks to category theory, many pairs of results/theorems in algebra
can be obtained with half the effort.

\subsection{Functors%
  \label{functors}%
}

The observation that there are some common features between different
categories intuitively brings the idea to transform one category into another,
while preserving its structure.

For example, our category of “pipes” over Unix data streams can be
transformed into a category of “networked services” over network data
streams trivially, by attaching the program \texttt{nc} around each Unix processes.

Such a transformation of a category into another is called a
\textbf{functor}. Generally, a functor \emph{F} from a category \emph{C} to a
category \emph{D} has the following properties:
\begin{itemize}

\item for each object \emph{x} in \emph{C}, \emph{F} associates an object in \emph{D} noted \emph{F(x)};

\item for each morphism $f : x\rightarrow y$ in C, \emph{F} associates a
morphism in \emph{D} that can be described by $F(x)\rightarrow
F(y)$, noted \emph{F(f)};

\item for every identity morphism $h : x\rightarrow x$ in \emph{C}, \emph{F(h)} is
an identity morphism for \emph{F(x)} in \emph{D};

\item for every pair of morphisms \emph{f} and \emph{g} in \emph{C}, $F(g\cdot f) = F(g)\cdot F(f)$ in \emph{D}.

\end{itemize}

(This specific flavor of functors is called “covariant.” If \emph{F} maps
each arrow in \emph{C} to an arrow with opposite direction in \emph{D}, it is
called “contravariant” instead. Covariant and contravariant functors are dual.)

\textbf{Side property:} Because of the properties of functors, the algebraic structure formed by 1) a set
of categories 2) a set of functors over these categories 3) the
identity functors that leave each category in set \#1 unchanged, and 4) the natural
generalization of composition, together, is itself a category.

\subsection{Some follow-up concepts to read about%
  \label{some-follow-up-concepts-to-read-about}%
}
\begin{itemize}

\item \textbf{Natural transformations:} a constuction that transforms
a functor into another functor, that respects the category
structure of the functor transformations.

\item \textbf{Functor category:} category where the objects are functors,
and the morphisms are natural transformations between them.

\item \textbf{Categorical logic}, especially used in computer science, focuses
on semantic systems with a difference between syntax and
semantics. To use categorical logic, one defines one category for
syntax, one category for semantics, and phrases interpretation as a
functor between them. An example application is proofs of behavior
correctness, or “correct by construction” languages: by choosing an
appropriate interpretation functor, proofs over the syntax category
can be carried over transparently to the category of semantics.

\end{itemize}

\section{Further reading%
  \label{further-reading}%
}
\begin{itemize}

\item Benjamin L. Russell. \href{https://dekudekuplex.wordpress.com/2009/01/19/motivating-learning-category-theory-for-non-mathematicians/}{Motivating Category Theory for Haskell for
Non-mathematicians}. Benjamin's Adventures in Programming Language
Theory Wonderland, January 2009.

\item James Cheney. \href{http://homepages.inf.ed.ac.uk/jcheney/presentations/ct4d1.pdf}{Category theory for dummies (I)}. Programming
Languages Discussion Group, March 2004.

\item José Antonio Ortega Ruiz. \href{http://programming-musings.org/2006/03/17/programmers-go-bananas/}{Programmers go banana}. Programming
musings, March 2006. Contains an extremely synthetic yet
approachable definition of categories with diagrams.

\item Gabriel Gonzalez. \href{http://www.haskellforall.com/2014/04/model-view-controller-haskell-style.html}{Model-view-controller in Haskell}. Haskell for
All, April 2014. Explains how to use category theory to abstract the
model-view-controller pattern of software engineering in a type-safe
manner.

\end{itemize}

\section{Acknowledgements%
  \label{acknowledgements}%
}

I am grateful to Karst Koymans for helping develop my
understanding of categories and denouncing major problems with an
earlier version of this document.

\DUtransition

\section{Copyright and licensing%
  \label{copyright-and-licensing}%
}

Copyright © 2014, Raphael ‘kena’ Poss.
Permission is granted to distribute, reuse and modify this document
according to the terms of the Creative Commons Attribution-ShareAlike
4.0 International License.  To view a copy of this license, visit
\url{http://creativecommons.org/licenses/by-sa/4.0/}.

\DUtransition

\href{http://www.structured-commons.org}{SC} fingerprint: \texttt{fp:boaT01NNxsK5g1kZfT\_03hFFhdOCwUzw4NjT1mCwBbIl2g}

\end{document}